\documentclass[a4paper]{article}

\usepackage{INTERSPEECH2020}

\title{Towards the Development of a Real-Time Deepfake Audio Detection System in Communication Platforms}
\name{Jonat John Mathew$^{1}$,
       Rakin Ahsan$^{1}$, Sae Furukawa$^{1}$,
       Jagdish Gautham Krishna Kumar$^{1}$,
       Huzaifa Pallan$^{1}$,
       Agamjeet Singh Padda$^{1}$,
       Sara Adamski$^{1}$,
       Madhu Reddiboina$^{1}$,
       Arjun Pankajakshan$^{1}$
       }
\address{$^1$ RediMinds Research, USA}
\email{\{firstname\}.\{lastname\}@rediminds.com}

\begin{document}

\maketitle
\begin{abstract}
Deepfake audio poses a rising threat in communication platforms, necessitating real-time detection for audio stream integrity. Unlike traditional non-real-time approaches, this study assesses the viability of employing static deepfake audio detection models in real-time communication platforms. An executable software is developed for cross-platform compatibility, enabling real-time execution. Two deepfake audio detection models based on Resnet and LCNN architectures are implemented using the ASVspoof $2019$ dataset, achieving benchmark performances compared to ASVspoof $2019$ challenge baselines. The study proposes strategies and frameworks for enhancing these models, paving the way for real-time deepfake audio detection in communication platforms. This work contributes to the advancement of audio stream security, ensuring robust detection capabilities in dynamic, real-time communication scenarios.
\end{abstract}


\section{Introduction}
Deepfake audio detection constitutes a crucial task, wherein the objective is to discern authentic (real/bonafide) audio samples from counterfeit (fake/spoof) ones through machine learning methodologies \cite{almutairi2022review, lyu2020deepfake}. Given the progress in text-to-speech (TTS) \cite{kaur2023conventional} and voice cloning (VC) \cite{arik2018neural} technologies, particularly in the context of voice-centric applications, it becomes essential to address the potential threats posed by audio deepfake attacks. This necessitates the implementation of effective countermeasures to ensure the security of services in such applications. The predominant focus of studies in deepfake audio detection has revolved around the realm of automatic speaker verification (ASV) systems \cite{wu2015asvspoof}. Notably, the ASVspoof challenge \cite{wu2015asvspoof, kinnunen2017asvspoof, todisco2019asvspoof, yamagishi2021asvspoof} has served as a catalyst for numerous advanced research endeavors in this domain.

Despite the promising strides in recent developments pertaining to deepfake audio detection and countermeasure strategies, it is noteworthy that a significant portion of these solutions has been conceived and assessed within the confines of static audio recordings. Here, the term \emph{static audio recordings} refers to datasets comprising audio samples lasting between $2$-$10$ seconds, characterized by limited variations in acoustic backgrounds, the number of speakers, encountered artifacts and recording conditions \cite{yamagishi2019vctk, ljspeech17}. In this work, we denote \emph{static deepfake audio models} to represent models trained using static audio recordings. It is important to acknowledge that these static deepfake audio models may not consistently exhibit robust performance when deployed in real-time scenarios, such as continuous audio streams coming from applications on a source computer device or similar scenarios in communication platforms. Consider, for instance, a Teams group call within an organization, which represents the latter case. The observed decline in the performance of static deepfake models during such events can primarily be attributed to their lack of awareness regarding dynamic variations inherent in real-time conversational speech data within communication platforms.

In this paper, the viability of employing static deepfake audio detection models in real-time and continuous conversational speech scenarios across a communication platform is systematically assessed. Two benchmark models are initially developed, leveraging Resnet and LCNN architectures as per \cite{alzantot2019deep} and \cite{nautsch2021asvspoof} respectively. Moreover, we developed an executable software application compatible with diverse operating system environments, streamlining the deployment of our deepfake audio detection systems. For real-time testing of our models in a communication platform, a new dataset was curated using actual Teams meeting sessions. To enhance the resilience of these models for real-time audio streams within a communication platform, diverse strategies are proposed as part of future works. The subsequent sections describe the methodology (Section \ref{sec:mtd}), dataset and experimental details (Section \ref{sec:data}), the deepfake audio detection models (Section \ref{sec:model}), the software platform (Section \ref{sec:soft}), and the results, discussion, and future work strategies (Section \ref{sec:result}).

\section{Methodology}
\label{sec:mtd}
\begin{figure}[t]
  \centering
\includegraphics[width=\linewidth,height=4.5cm,keepaspectratio]{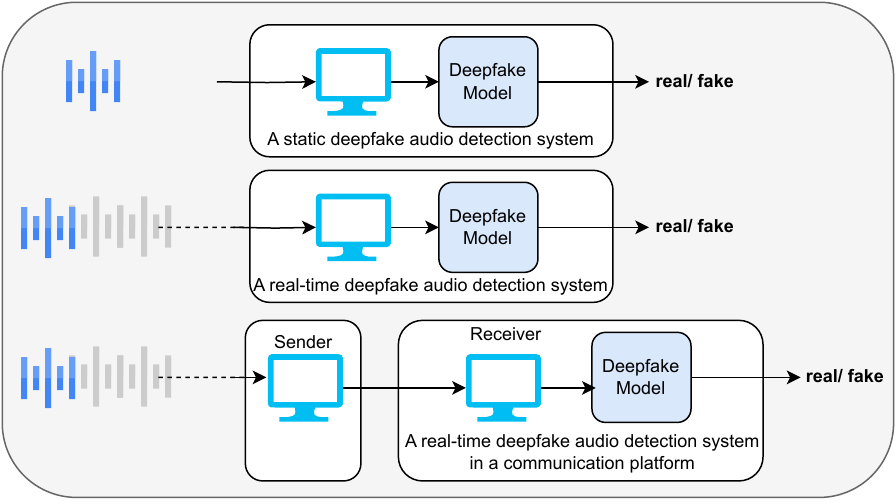}
  \caption{A block diagram of various deepfake audio detection systems.}
  \label{fig:bk_df}
\end{figure}

A block diagram illustrating the components of a static deepfake audio detection system, a real-time deepfake audio detection system implemented on a source device, and a real-time deepfake detection system within a communication platform is depicted in Figure \ref{fig:bk_df}. This study focuses on assessing and exploring the application of static deepfake audio detection models within a real-time audio communication platform. The methodology employed in this study is summarized as follows.

\begin{figure}[t]
  \centering
\includegraphics[width=8cm,height=5cm]{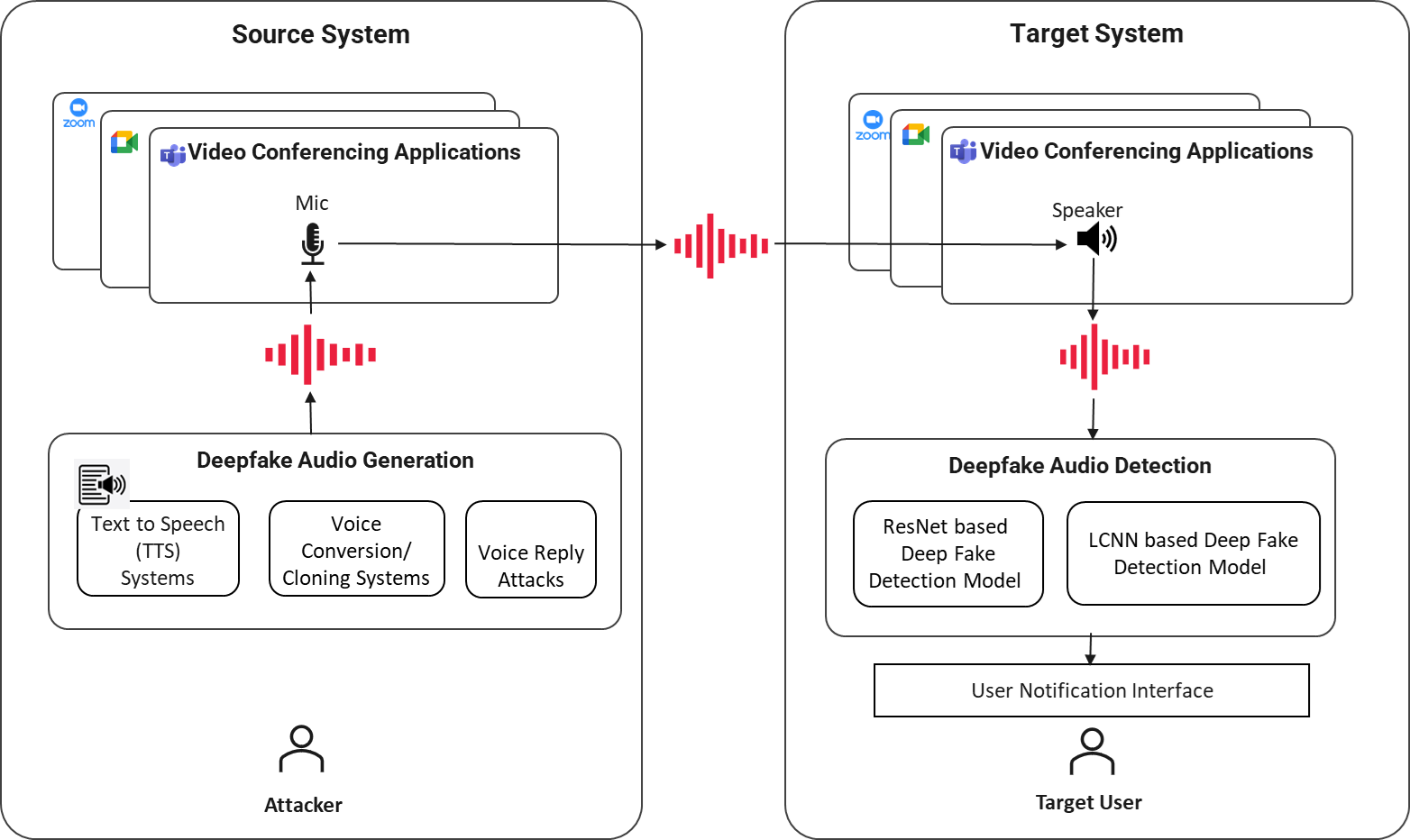}
  \caption{A block diagram showing the typical use-case of our deepfake audio detection system.}
  \label{fig:system_bk}
\end{figure}

\begin{itemize}
    \item Two static deepfake detection models are initially implemented, utilizing Resnet \cite{alzantot2019deep} and LCNN \cite{nautsch2021asvspoof} architectures. These models are trained on the ASVspoof $2019$ challenge dataset \cite{todisco2019asvspoof}.
    \item The benchmarking of the deepfake models is conducted by comparing their performance against the baseline model scores from the ASVspoof $2019$ challenge \cite{todisco2019asvspoof}.
    \item An executable software platform is developed to facilitate the deployment of the deepfake audio detection model. The software is designed to be compatible with various operating systems.
    \item Evaluation of the static deepfake models in real-time detection scenarios is carried out during our daily group meeting sessions on Microsoft Teams. This assessment involves the analysis of the model's performance in differentiating between real and fake voices in audio stream data over a communication platform. A block diagram view of the entire system is depicted in Figure \ref{fig:system_bk}.
\end{itemize}

This study aims to analyze the challenges and shortcomings associated with employing a static deepfake audio model for real-time deepfake audio detection on audio stream data within a communication platform. Building upon the insights gained from this analysis, we put forth several strategies to implement an effective real-time deepfake audio detection model tailored for deployment in a communication platform.

\section{Dataset and Experimental Details}
\label{sec:data}
The implementation of our deepfake audio detection models relies on the ASVspoof $2019$ challenge dataset \cite{todisco2019asvspoof}, which encompasses both logical access (LA) and physical access (PA) challenges. In the context of voice-based LA attacks, attackers seek to trick a system by imitating or altering an individual's voice, aiming to gain unauthorized digital access, such as unlocking a smartphone or accessing secure online information. These attacks are commonly executed through synthetic voice generation methods, including text-to-speech (TTS), voice cloning, and voice conversion technologies. The LA challenge is designed to distinguish between genuine voices and synthetic voices. On the other hand, voice-based PA attacks involve attackers to manipulate or deceive a voice-based biometric system by replaying a recorded real voice of an individual. The objective of the PA challenge is to distinguish between authentic voices and replayed voices.

The challenge comes with distinct LA and PA datasets, each comprising $8$ male and $12$ female speakers in both training and development subsets. In the LA challenge, the training subset encompasses $2580$ bonafide and $22800$ spoofed utterances, while the PA challenge includes $5400$ bonafide and $48600$ spoofed utterances. Evaluation sets for LA and PA consist of approximately $80000$ and $135000$ test utterances, respectively.

The initial phase involves the implementation of a Resnet model, adapted from \cite{alzantot2019deep}, to distinguish between real and synthetic voices. This model is exclusively trained using the LA dataset. Similarly, an LCNN model, adapted from \cite{nautsch2021asvspoof}, is implemented to distinguish between real and replayed voices, utilizing only the PA dataset. Subsequently, the performance of these models is assessed using the evaluation subset of the PA and LA datasets. The evaluation metrics employed for benchmarking include the tandem detection cost function (t-DCF) \cite{kinnunen2018t} and the equal error rate (EER) \cite{cheng2004method}. Lower scores in these metrics indicate better model performance.

The evaluation of our models focuses primarily on the ASVspoof $2019$ evaluation data. For real-time testing in a communication platform, we use Microsoft Teams group meeting sessions and create a new dataset specifically for this purpose, known as the \emph{Teams Meeting dataset}. 

The dataset is generated through the following procedure: we develop an executable software application capable of running our deepfake audio detection models. During Teams meeting sessions, our application is run on a participant's device, with another participant speaking while keeping their cameras turned off. This arrangement ensures the collection of authentic voice samples for testing the model predictions. To introduce synthetic voice events, we used voice cloning and voice conversion systems. A moderator is assigned to instruct a speaker to conduct either a real or a fake trial, allowing us to note the reference annotations associated with each trial. The topics for the trials include reading exercises and conversational speech between two speakers on selected topics. Each trial of a speaker lasts for $10$ seconds, and a total of $10$ individuals participate in the experiment, each contributing $5$ real and fake voice events from the reading exercises. The number of instances created using the conversational speech setup for real and fake scenarios is not equal.

To avoid scenarios where multiple speakers speak simultaneously (cocktail party scenario), we organized individuals participation in the experiment carefully with proper time-slot arrangement. Throughout the $10$-second voice trail, model predictions are aggregated, and a majority voting scheme is employed to classify the event as either real or fake. This approach enables us to evaluate the individual accuracy of our static LA model (ResNet model) and PA model (LCNN model) for predicting real-time audio stream data over a communication platform. 

\section{Model pipeline and Training}
\label{sec:model}
\begin{figure}[t]
  \centering  \includegraphics[width=\linewidth,height=4cm,keepaspectratio]{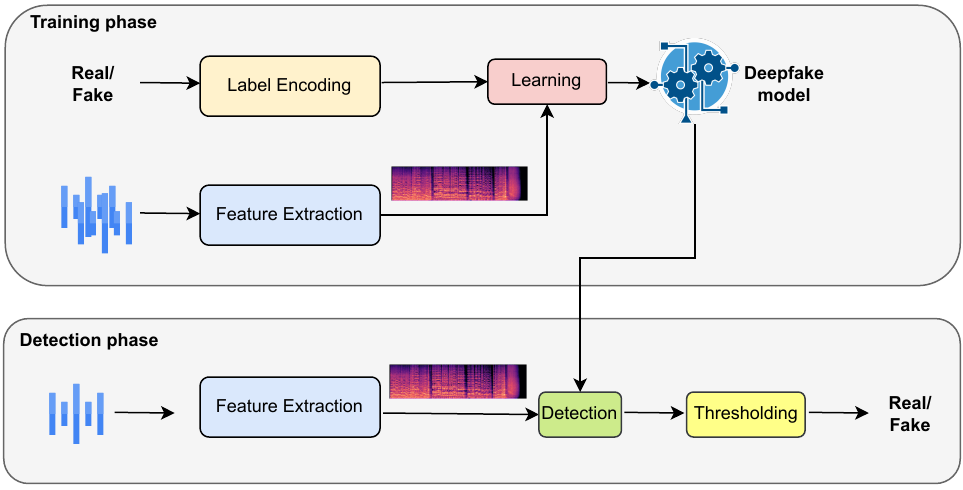}
  \caption{Pipeline of a deepfake audio detection system.}
  \label{fig:pipeline_df}
\end{figure}

The developmental and evaluation stages of our deepfake audio detection systems follow a sequential pipeline, as depicted in Figure \ref{fig:pipeline_df}. The training and development phase initiates with the construction of an acoustic model, which learns a mapping function between audio samples and their reference annotations. Raw audio samples undergo preprocessing in the feature extraction unit, extracting compressed and information-rich audio feature representations. Several audio feature representations, such as mel-spectrogram cepstral coefficient (MFCC), power-spectrogram, and mel-spectrogram features, are empirically experimented with for the development of our deepfake audio model. Ultimately, mel-spectrogram features are selected for the LA model, while power-spectrogram features are chosen for the PA model. The label encoding unit transforms textual reference annotations (real/fake) into a numeric format, utilizing binary encoding as the label encoding function. In the detection phase, a threshold value, corresponding to the minimum EER computed using the validation dataset, is applied to the probability score from the model. This threshold aids in identifying whether a given audio sample is classified as real or fake.

Our LA model is constructed using an adapted version of the Resnet model architecture from \cite{alzantot2019deep}, while our PA model employs an adapted version of the LCNN model architecture as presented in \cite{nautsch2021asvspoof}. Training our models involves utilizing audio samples of a $3$-second duration. To standardize input lengths, trimming and padding techniques are applied to long and short audio samples in the LA and PA datasets, resulting in $3$-second audio samples. For the LA model, a feature representation of $80$ log mel-bands spectrogram is employed. This feature is extracted using a short-term Fourier transform (STFT) with a fast Fourier transform (FFT) window of $512$, a hop length of $160$, and a sample rate of $16$ kHz. On the other hand, the PA model uses a power-spectrogram feature representation, extracted with the same feature settings as above, but without the mel-transformation.

For both the PA and LA models, each convolutional layer activation undergoes batch normalization and is subject to regularization with dropout (probability = $0.3$). The weights of the convolutional layers are initialized using a Glorot uniform distribution \cite{glorot2010understanding}. Training for each model spans $100$ epochs, employing a binary cross-entropy loss function, as defined in equation (\ref{eq:bce}), and the Adam optimizer with a learning rate set at $0.001$. To mitigate overfitting, early stopping is implemented, with termination set after $10$ epochs based on the validation loss score.

\begin{equation}
\begin{split}
  \mathcal{L} &= -\sum_{n=1}^N{(y_n\log(p_n) + (1 - y_n)\log(1 - p_n))}
  \label{eq:bce}
\end{split}
\end{equation}

In Equation (\ref{eq:bce}) $\mathcal{L}$ is the binary cross-entropy loss function, $N$ is the total number of audio samples in the training dataset and $y_n$ and $p_n$ respectively denotes the true label and predicted probability for the $n^{th}$ audio sample.

\section{Software application development}
\label{sec:soft}
\begin{figure}[t]
  \centering
  \includegraphics[width=\linewidth,height=12cm,keepaspectratio]{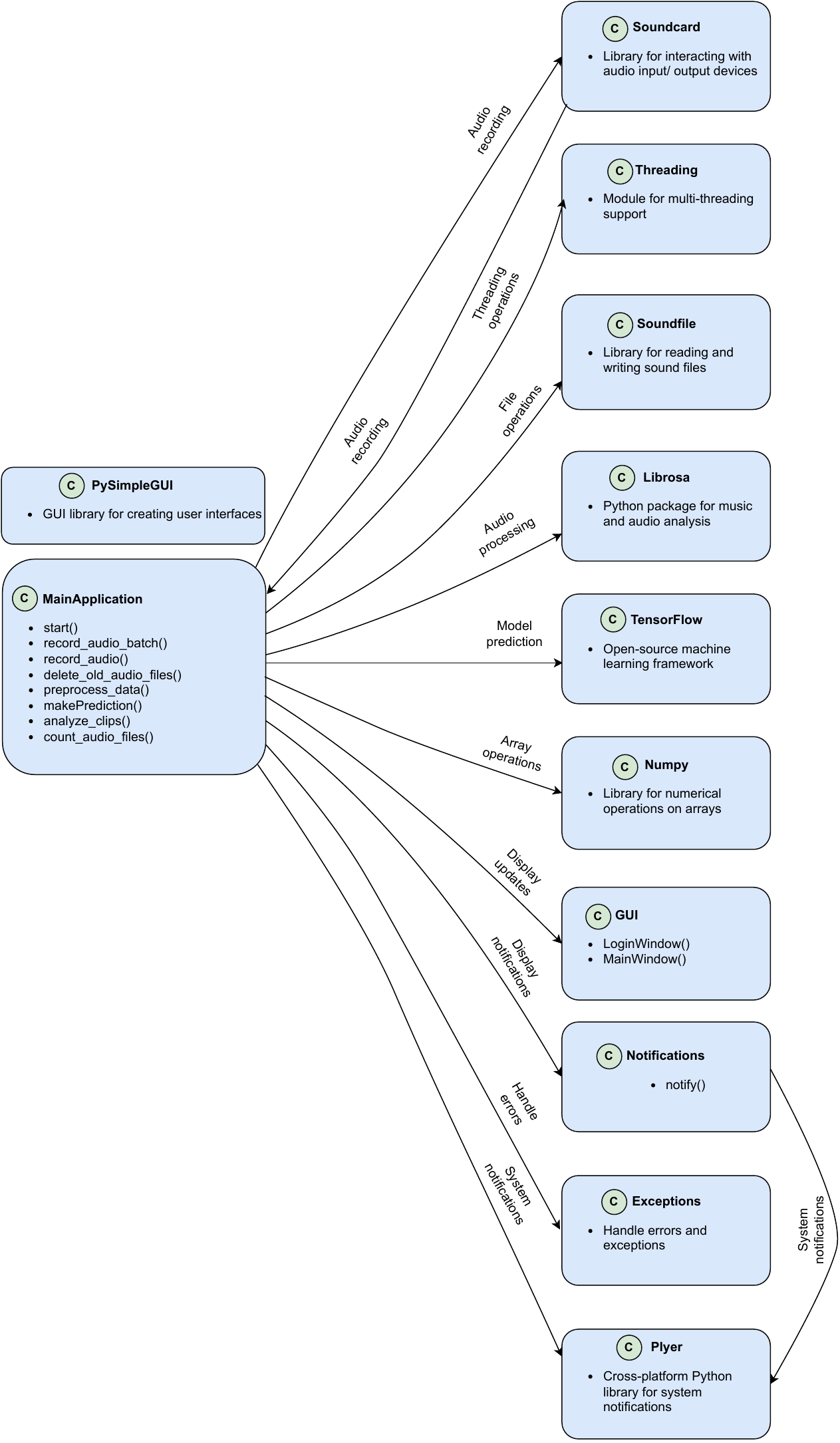}
  \caption{Flow diagram of the software application development process.}
  \label{fig:soft}
\end{figure}

We have developed a software application designed for use alongside popular conferencing platforms such as Teams and Zoom, aimed at enhancing the integrity of audio interactions during remote communications. Python serves as the chosen programming language for this initiative, selected for its versatility. The ecosystem of libraries and frameworks in Python provides an ideal environment for implementing AI and advanced audio processing techniques. Our program architecture adheres to the MoSCoW rules, ensuring a selection of core functionalities. A flow diagram of the complete development process in shown in Figure \ref{fig:soft}.

The graphical user interface (GUI) is designed, incorporating a secure login mechanism and a multifunctional main screen. On the backend, the program orchestrates the automated recording of audio from default speakers, real-time analysis of audio clips, and the utilization of an AI model for classification of audio authenticity. The key functionalities of our application are the following.

\begin{itemize}
    \item GUI Interface: A user interface crafted for intuitive interaction, encompassing a secure login screen and a main interface with essential functionalities.
    \item Automated Audio Recording: Designed to autonomously capture audio clips from default speakers in real-time during a video conference session. 
    \item AI Enabled Deepfake Audio Detection Models: The pinnacle of our technological innovation, an intricately crafted AI model leverages advanced techniques to discern between authentic and fake audio streams. 
    \item Real-time Notifications: An Operation System level  notification system to promptly inform the user of the results. 
    \item Autonomous Operation: Equipped with automated processes, the program consistently maintains its operations after initiation, ensuring continuous functionality. 
\end{itemize}

As illustrated in Figure \ref{fig:soft}, the codebase of our software program is distinguished by structuring into modular components. This flow diagram shows the architecture of the Python application that involves audio recording, processing, and system notifications. It uses various libraries such as PySimpleGUI, Soundcard, Librosa, Plyer, and TensorFlow to perform different tasks within the application. The diagram illustrates the main components and methods of the \emph{MainApplication class}, as well as the interactions and dependencies among the libraries. 

Each module serves a specific purpose, encapsulating functionalities for audio recording, preprocessing, prediction, and GUI interactions. Threading intricacies have been employed to enable parallel execution, aligning with industry-best practices for code readability and maintainability. This structure enhances the clarity of the codebase and also promotes scalability and ease of future development. We test our system in stages through trial and error, checking each part to confirm it works correctly, including testing on different platforms.

\section{Results and Discussion}
\label{sec:result}
In this section, experimental results are presented and analyzed using the ASVspoof $2019$ evaluation data and our Teams Meeting dataset. It is crucial to note that our models are trained solely using the ASVspoof $2019$ development dataset, and the Teams Meeting dataset is exclusively used for real-time testing of our models.

\subsection{ASVspoof 2019 Evaluation data}
The deepfake audio detection performance of our LA and PA models on the ASVspoof $2019$ evaluation data is summarized in Table \ref{table:LA-PA-ASV}. The baseline models, denoted as Baseline 1 and Baseline 2 in Table \ref{table:LA-PA-ASV}, correspond to the baseline models associated with the ASVspoof $2019$ challenge \cite{todisco2019asvspoof}. The results indicate that our LA model achieves an EER of $7.39$, outperforming both baseline models. Additionally, the LA model demonstrates a t-DCF score of $0.215$, closely aligning with the score of the Baseline 1 model. Similarly, our PA model exhibits an EER of $4.38$ and a t-DCF score of $0.121$, surpassing the baseline model scores. These findings underscore the effectiveness of our static PA and LA models in distinguishing between real and fake voices within the context of the ASVspoof $2019$ dataset.

\subsection{Teams Meeting dataset}
Table \ref{table:LA-PA-teams} provides an overview of the performance metrics for the LA and PA models when deployed in real-time audio stream data during Microsoft Teams meetings. The metrics for evaluation include precision (P), recall (R), and F-score (F), calculated based on the definitions provided in the equation (\ref{eq:p-r-f}), where $TP$, $FP$, and $FN$ represent true positive, false positive, and false negative rates, respectively. The precision values indicate the proportion of correctly identified positive instances among all instances predicted as positive. In this context, the LA model outperforms the PA model with a higher precision of $0.48$ compared to $0.39$, suggesting a better ability to correctly identify positive instances. Similarly looking at the recall scores, it can be identified that the PA model is more prone to false negative predictions.

\begin{equation}
    P = \frac{TP}{TP+FP}, \quad R = \frac{TP}{TP+FN}, \quad F = \frac{2PR}{P+R} \\
    \label{eq:p-r-f}
\end{equation}

The results reveal that the models exhibit poor performance in real-time communication platform scenarios compared to their performance on the ASVspoof $2019$ evaluation data. This underperformance can be attributed to the static nature of the audio data used for training the models, which contrasts with the dynamic nature of real-time audio stream data. The ASVspoof $2019$ dataset's training and development subsets share similar spoofing algorithms and conditions for both the LA and PA datasets. We assume that training on the same types of spoofing attacks may lead to overfitting and limited generalization on unseen attack conditions. Additionally, the dataset's limited variations in terms of the number of speakers, acoustic conditions, and recording conditions contribute to poor generalization on real-time audio stream data.

To address these issues and enhance the robustness of our deepfake audio detection models for real-time communication platforms, we propose the following strategies.

\subsection{Recommended Strategies}

\subsubsection{Training Data Variability}
To address real-time communication nuances, augment the ASVspoof $2019$ dataset with additional conversational, noisy, and fragmented speech data is an immediate first step. Incorporating datasets like the \emph{Spotify podcast} collection \cite{clifton2020100}, \emph{FoR} \cite{reimao2019dataset}, \emph{WaveFake} \cite{frank2021wavefake}, \emph{LRPD} \cite{yakovlev2022lrpd}, \emph{In-the-wild} \cite{zi2020wilddeepfake}, and \emph{VoxCeleb2} \cite{chung2018voxceleb2} can introduce more diverse acoustic scenarios, enhancing the model's adaptability to real-world communication platform conditions. We plan to investigate the effects of training with these datasets collectively in our future work.
\subsubsection{Data Augmentation}
Develop novel data augmentation methods tailored to the specific variations introduced by communication platforms. This may involve simulating platform-specific artifacts, such as network-induced delays, packet loss, or compression artifacts. These techniques aim to improve the model's robustness to the unique challenges posed by real-time audio streams in communication environments. In our future work, we plan to explore a data augmentation strategy that focuses on understanding the impact of data compression in the communication platform.
\subsubsection{Generative Model Approaches}
Explore the integration of generative models to synthesize additional training samples. By leveraging generative models like variational autoencoder (VAE) \cite{kingma2019introduction} or similar approaches, synthetic data can be generated, providing the model with a more extensive set of examples for distinguishing between genuine and deepfake audio in real-time scenarios. In our future work, we plan to design and implement a generative model-augmented classifier for real-time deepfake detection, inspired by the approach proposed in \cite{chettri2020deep}.

\begin{table}[t]
\centering
\renewcommand{\arraystretch}{2}
\caption{Evaluation performance of the LA and PA model using ASVspoof $2019$ evaluation dataset.}
\label{table:LA-PA-ASV}
\begin{tabular}{|l|ll|ll|}
\hline
\multicolumn{1}{|c|}{\multirow{2}{*}{\textbf{Model}}} & \multicolumn{2}{c|}{\textbf{LA}}                    & \multicolumn{2}{c|}{\textbf{PA}}                    \\ \cline{2-5} 
\multicolumn{1}{|c|}{}                                & \multicolumn{1}{l|}{EER}           & t-DCF          & \multicolumn{1}{l|}{EER}           & t-DCF          \\ \hline
Baseline 1                                            & \multicolumn{1}{l|}{8.09}          & \textbf{0.211} & \multicolumn{1}{l|}{13.54}         & 0.301          \\ \hline
Baseline 2                                            & \multicolumn{1}{l|}{9.57}          & 0.236          & \multicolumn{1}{l|}{11.04}         & 0.245          \\ \hline
LA                                                 & \multicolumn{1}{l|}{\textbf{7.39}} & 0.215          & \multicolumn{1}{l|}{-}             & -              \\ \hline
PA                                                 & \multicolumn{1}{l|}{-}             & -              & \multicolumn{1}{l|}{\textbf{4.38}} & \textbf{0.121} \\ \hline

\end{tabular}
\end{table}

\begin{table}[t]
\centering
\renewcommand{\arraystretch}{2}
\caption{Evaluation performance of the LA and PA model on real-time audio stream data in Teams.}
\label{table:LA-PA-teams}
\begin{tabular}{|c|l|l|l|}
\hline
\textbf{Model}   & Precision & Recall & $F$-score \\ \hline
\textbf{PA} &    0.39       &    0.41    & 0.40 \\ \hline
\textbf{LA} &      0.48     &    0.42    &    0.45       \\ \hline
\end{tabular}
\end{table}

\section{Conclusion}
This study aimed to develop a real-time deepfake audio detection system for communication platforms. Two static deepfake audio detection models were implemented using the ASVspoof $2019$ dataset. Subsequently, a software application was created to assess the models performance in real-time communication, specifically tested in Teams meeting sessions. The investigation highlighted challenges associated with deploying static deepfake detection models in real-time communication platforms. The study recommends future work to focus on developing an efficient deepfake audio detection model for real-time prediction in communication platforms. 

\bibliographystyle{IEEEtran}

\bibliography{mybib}

\end{document}